# Photoinduced ultrafast dynamics and control of chemical reactions: from quantum to classical dynamics

Leticia González and Philipp Marquetand
Institute of Theoretical Chemistry, University of Vienna,
Währinger Straße 17, 1090 Vienna, Austria

leticia.gonzalez@univie.ac.at

## 1. Introduction

A challenging field of research is to develop tools that are able to follow the nuclear motion in chemical reactions, e.g. when a molecule is breaking and forming chemical bonds or simply reorganizing its atoms to form a different isomer. As shown by the pioneering experiments carried out by Zewail and coworkers (see e.g. [1]) nuclear motion occurs on the ultrafast time scale of femtoseconds (fs). Therefore, it was only with the advent of fs lasers that the dream to observe elementary processes in real time was born. Furthermore, the availability of more and more complex shaped laser pulses has permitted manipulating molecular motion at wish.

Since its birth, the field laser control of chemical reactions, which can be understood as using laser pulses to transform reactants into products with the maximum efficiency, has gained a lot of adepts –as illustrated by the number of excellent reviews which can be nowadays found on this field [2,3,4,5,6,7,8,9,10,11,12,13,14,15,16,17].

From the practical point of view, the dream of controlling the outcome of a reaction is not only justified because it offers a possibility to maximize (or minimize) a particular product and minimize (or maximize) a by-product. It is also attractive because it opens the door to the synthesis of novel molecular species, the design of new materials, the execution of chemical and biological functions, and even steering new physical phenomena using a clean and efficient form of energy. This dream is also our dream and our group has made numerous contributions to the field, which will be given below.

Control is a well-known concept in traditional Chemistry. Control of chemical reactions can be achieved thermodynamically or kinetically. Thermodynamic control relies on modifying external factors, like temperature, pressure or concentration, in order to modify the reaction equilibrium. Kinetic control is based on introducing a catalyst to reduce the energy of a transition state barrier, therefore favoring a particular reaction channel. These tools, however, do not access the microscopic behavior of a chemical reaction. For this reason, this type of control is often termed *passive control.* Light can also be used passively to modify thermodynamically or kinetically a chemical reaction. When a system is photo-excited from the ground to some electronic excited state, it is possible to change the relative free energy between reactants and products, thus changing thermodynamically the system. A well-chosen wavelength can excite the system above a particular reaction barrier and lead kinetically to a selected product. In any of these cases the success of this, one would called it, "traditional photochemistry" relies on the quantum mechanical temporal evolution of the system, which in turn depends on the profile of the multidimensional potential energy surfaces (PES). As with temperature, pressure or any other external macroscopic variables, light used in this way does not access the microscopic behavior of a chemical reaction. Consequently, it is adventurous to predict the outcome of a reaction after plain irradiation, let alone to control it.

When nanosecond (ns) lasers became available in the 1960s, it was expected that reaction paths could be steered in a more controlled way. The underlying idea was that tuning a laser with the vibrational frequency of a particular bond would weaken the bond and finally break it, leading to a particular product. Unfortunately, this concept, so-called *mode-selective chemistry*, ran up against an unexpected difficulty: the different degrees of freedom are –in most cases– coupled to each other and after a short time the energy deposited in a particular bond was quickly redistributed through the whole molecule and unluckily control could not be achieved. The problem was that the pulse duration of ns lasers was much too long to avoid internal energy vibrational redistribution in all, but exceptional cases. What now it is known as *active control* over the nuclear dynamics requires fs resolution and had to wait until the development of Femtochemistry (see also Ref. 18). This article is devoted to review some of the most standard ways to control chemical reactions using laser pulses, putting special emphasis on the applications that are closer to our research interests.

## 2. Schemes for ultrafast laser control of chemical reactions

Theoretically, it was early recognized that matter can be manipulated with light if the coherence properties of the quantum mechanical wave function are exploited. This was known as *coherent control*, even when today this term is often used more loosely to designate any control achieved with coherent light, i.e. with laser light.

In general, laser control can be achieved changing the spectral properties of the coherent light so that an initial quantum mechanical wavefunction is transformed into a final desired one. This transformation can be achieved in different ways. If only one parameter of the laser is changed at a time, a possible and common classification of control schemes is: a) control in the frequency domain, b) control in the temporal domain, and c) adiabatic control. In addition, it is also possible to control chemical reactions using d) the chirp or e) strong fields. Despite a large amount of experimental success (see e.g. Refs. [19,20,21]), in most but the simplest systems, the search for an appropriate chemical path is hampered by the incomplete knowledge of the molecular Hamiltonian. For this reason, techniques have emerged, which - using learning algorithms - are able to optimize in an iterative manner the many laser parameters that are necessary to control simultaneously all the coupled degrees of freedom. This type of control will be explained below under section f) multi-parameter laser control scheme.

a) Control in the frequency domain

To obtain control in the frequency domain, as proposed by Brumer and Shapiro [22,23], the initial state is prepared as a coherent superposition of bound states with a well-defined phase. Constructive and destructive interferences between two independent excitation routes are exploited, leading to the enhancement or prevention of a particular reaction channel. Experimentally, this type of control requires two different monochromatic excitation sources, providing e.g. 3 photons with frequency $\omega$ or a photon with frequency $3\omega$ (see Fig. 1), as originally proposed in Ref. [24].

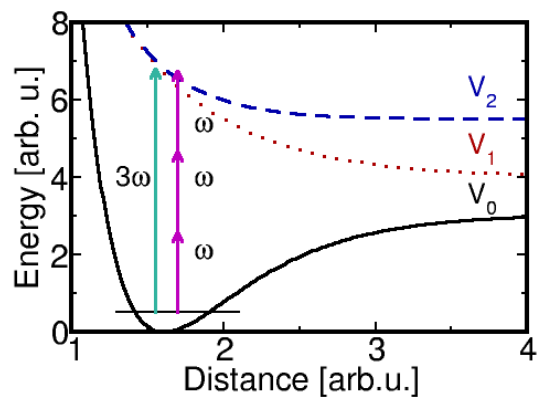

**Fig 1: Control in the frequency domain (Brumer-Shapiro scheme). Degenerate states are coupled by two laser pulses of different frequency. Constructive and destructive interference between both pathways leads to the desired outcome.**

The experiment is related to the double slide experiment of Young for particles or beams of light: An initial molecular state can follow two different pathways characterized by two different wave functions, say $\psi_1$ and $\psi_2$, on the initially degenerate potentials $V_1$ and $V_2$. The probability of the final event is proportional to the square of the sum of the quantum mechanical amplitudes associated to each of the independent pathways connecting the initial and final state. In the case of the two coherent light beams, one obtains interference patterns of enhanced or reduced probabilities. In the case of coherent control, the overall coherence of the state and the light source allows for a constructive or destructive manipulation of probabilities. Since the final wave function depends on the difference of phases of the respective $\psi_1$ and $\psi_2$, by modulating the individual transitions between initial and final state, it is possible to modulate the probability of formation of a particular state. An early example was the control of the branching ratio in the dissociation of IBr: IBr→ I+Br or I+Br*, where the star indicates the spin-orbit excited state of the Br atom versus its ground state (denoted without star) [25]. Other early examples can be found in Refs. [26,27].

b) Control in the time domain

Control in the temporal domain was first suggested by Tannor, Kosloff and Rice [28,29], see also [30], and requires to control the time-delay between two ultrashort pulses, e.g. a pump and a dump. This strategy was the first one that exploited the inherent broad spectral band of an ultrashort fs laser pulse. In its original formulation, first a pump pulse interacts with a molecular system to create a linear combination of vibrational eigenfunctions (a wave packet) in an electronically excited PES at the initial time $t_0$ (see Fig. 1B). This wave packet will evolve in time according to the topology of the excited PES. Quantum mechanically, it changes its composition of the vibrational eigenstates; classically, this corresponds to a change on the vibrations of some atoms in the system. If the excited state is bound around the Franck-Condon region, the wave packet will be trapped giving rise to some oscillations until ultimately delocalizes due to the anharmonicities of the potential.

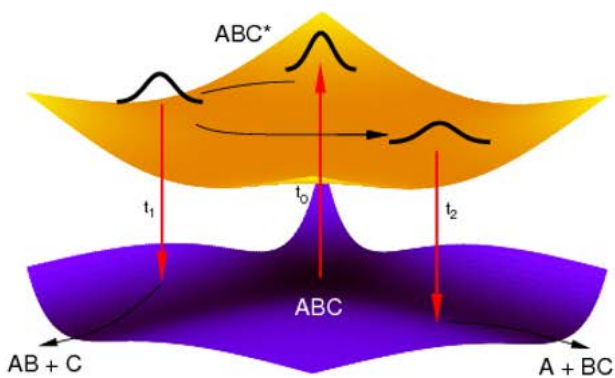
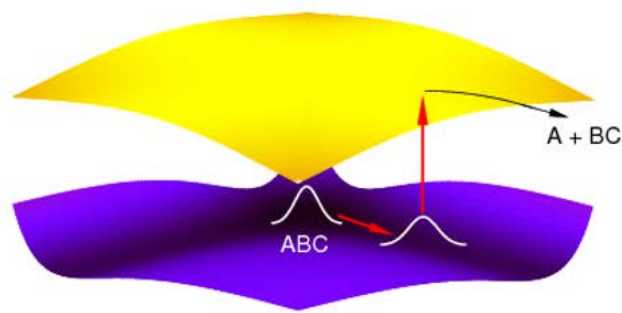

A) Tannor-Kosloff-Rice scheme

B) IR+UV scheme

**Fig. 2: Control in the time domain. A) Tannor-Kosloff-Rice scheme. A molecule ABC is electronically excited by a first (pump) pulse and the control is exerted by correctly timing a second (dump) pulse. B) IR + UV scheme. A few-cycle IR laser pulse creates a vibrational wavepacket in the electronic ground state. A subsequent timed UV laser pulse regime transfers population from the ground to the excited state potential.**

If on the contrary, the excited state is unbound, the wave packet will quickly delocalize evolving in the direction of steepest decent. Monitoring either dynamics requires a second pulse, which experimentally is typically called the probe pulse. Interestingly, depending on the topology of the excited PES, the wave packet visits different parts of the PES at different times. Thus, if the probe pulse dumps the wave packet back to the electronic ground state at particular times, say $t_1$ or $t_2$, particular molecular configurations can be obtained, leading to different products in the ground state; see Figure 2A. In general, however, the dump pulse can move the wave packet to any other suitable PES, and the final products can be also obtained in an excited PES.

Experimentally, the Tannor-Kosloff-Rice method has been effectively employed to control diatomic and triatomic systems, see the early examples of $Na_2$ into $Na_2^+ + e^-$ versus $Na^+ + Na + e^-$ [31,32] or the control over the yield of $XeI$ molecules from $Xe + I_2$ [33]. The extension of pump-dump control to larger systems is nevertheless not trivial. This type of control relies very much on exciting to the appropriate PES and finding the adequate time delay at which the wave packet should move from one PES to another. In large molecular systems, it is very difficult to know the topology of the multidimensional PES in advance and therefore it is not straightforward to predict the optimal wavelength and the optimal time delay between the pump and dump pulse. This is one of the reasons, why scans over the different laser parameters can be very useful (see also below).

There are different ways in which the pump-dump control strategies can be modified. For instance, it is not necessary that the first pump pulse creates a superposition of vibrational states in an electronic excited state. Such superposition, a wave packet, can also be created in the electronic ground state, e.g. after burning a hole, or with few-cycle laser pulses. Few-cycle pulses are pulses where the pulse duration is only a small multiple of an optical cycle. The use of few-cycle IR followed by an UV laser pulse has proofed useful to control chemical reactions. The underlying mechanism can be considered similar to that of the pump-dump control because the time delay between the IR and the UV is the key step to control the reaction. The control works in the following way. First, the few-cycle IR laser pulse is applied to the system creating a wave packet in the electronic ground state. The created wave packet evolves in the PES. When the wave packet is at the appropriate position, a UV pulse is fired and transfers probability density to the target PES, in which the reaction takes places along the desired channel (see Fig. 2B). It is important that the UV laser pulse is shorter than the vibrational period of the ground state wave packet; otherwise, the scheme is not efficient. The first application of the few-cycle IR+UV laser control was demonstrated in the theoretical selective bond dissociation of symmetric triatomic molecules: the isotopically substituted ozone $^{16}O^{16}O^{18}O$ [34], HOD [8,35,36], and the strong hydrogen bonded systems $FHF^-$ and $FDF^-$ [37,38,39]. Other applications in asymmetric molecules have followed, as e.g. in $OHF^-$ [40] or $CH_2BrCl$ [41].

In the examples above, control is accomplished by selecting the time delay between the IR and the UV laser pulses so that the wave packet is transferred when its position is the most adequate. It is also possible to control the time delay so that it is not the position but the momentum, which is optimal to achieve a particular chemical reaction. Control of momenta directed to selective bond breaking has been demonstrated in HOD [42]. In such a case one aims at creating momentum along the bond to be broken, so that the UV pulse transfers the wave packet to a different electronic state with momentum along a particular product channel. The UV pulse is not fired when the OH or OD bonds are maximally or minimally stretched, that is, at the turning point of the ground state potential but at the position of the equilibrium geometry, where momentum is maximal. After few oscillations, the wave packet has accumulated momentum and depending on the time delay, it is possible to control the branching ratio by directing this momentum.

This strategy can be also be wisely employed in other contexts, e.g. to ignite molecular rotors. In references 43,44, unidirectional torsional motion can be driven by few-cycle IR + UV pulses. As in the previous examples, the key step in the control is the creation of a wave packet in the electronic ground state with a few-cycle IR laser pulse. Here, the linearly polarized IR field creates torsional angular momentum in the electronic ground state, which will be transferred to the excited state by the UV field. In passing we note that molecular rotors have also been ignited using pump and dump pulses [45].

c) Adiabatic control

An adiabatic technique to control population transfer was introduced by Bergmann and coworkers [46,47] and named stimulated Raman adiabatic passage (STIRAP). STIRAP has a certain analogy to the Tannor-Kosloff-Rice method since it is based on using two lasers, which are fired at a convenient time delay, but it uses a counter-intuitive and overlapping sequence of pulses. In its original version, STIRAP was designed to achieve complete population transfer in a three-level lambda system, with two low levels initially populated and non-resonantly coupled to an initially empty upper level (see Fig. 3). The pump laser couples the initial state |1> with the intermediate state |2>; the Stokes laser couples the intermediate state |2> and the final state |3>. The employed laser fields are strong enough such that many Rabi oscillations are generated between |1> and |2> and between |2> and |3>. A Rabi oscillation is the cyclic behavior of a two-state quantum system in the presence of an oscillatory driving field. By overlapping counter intuitively the Stokes and pump pulses and satisfying adiabaticity requirements [48], it is possible to achieve a complete population transfer from the initial state |1> to the final one |3> without populating the intermediate state |2>. The advantage of this method is that because it avoids populating intermediate states and it is independent from dissipative process, which might take place at the intermediate state.

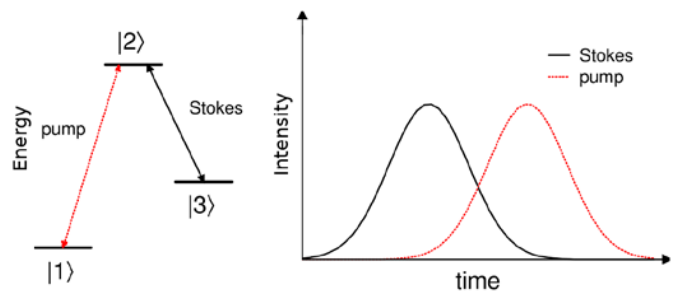

Fig. 3: With the STIRAP method, complete population transfer from initial state |1> to final state |3> is possible. The states are coupled by two overlapping nanosecond pulses (pump pulse and Stokes pulse), which are applied in a counter-intuitive way. The transient population of the intermediate state |2> remains zero.

The first experimental realization of STIRAP was designed to achieve population transfer between the excited states of Ne [49]. Beyond atoms, adiabatic strategies have been applied in small polyatomic systems, like $SO_2$ [50]. Theoretically, STIRAP has been proposed as a way to convert a racemic mixture into pure enantiomers [51] or to repair a base pair mutation [52], to mention few examples.

d) Strong field control

A different approach to control the outcome of a chemical reaction is to use strong resonant or non-resonant laser fields ($\approx 10^{13}$ Wcm$^{-2}$). The control with strong pulses, on the one hand, is based on the ability to induce the nonresonant dynamic Stark effect (NRDSE) which can displace the energetic levels and change the potential landscape for a reaction in the adequate direction. On the other hand, multiphoton transitions can be induced with strong fields (see Fig. 4). Very short pulses are often used to obtain the necessary intensities. Consequently, the bandwidth can be so large that even electronic wave packets can be created, where different electronic states are coherently excited [53]. A strong non-resonant pulse can modify potential energy barriers without inducing a real transition, and therefore act like a photonic catalyst.

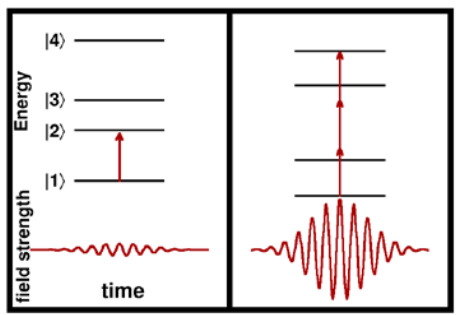

**Fig. 4: Multiphoton processes and NRDSE.** Strong laser fields (right panel) may shift the quantum states of a molecule as compared to a weak pulse (left panel). Moreover, multiphoton excitations are possible with strong fields while weak fields usually induce only single-photon transitions.

Non-resonant strong fields have been applied to control the population transfer in model polyatomic systems [54] and to control non-adiabatic processes [55,56]. An interesting example is shown in Ref. [55]. There it is shown that the photodissociation of IBr and thus the branching ratio between Br or Br* can be catalytically controlled by Stark-shifts. Extensions of this type of control for ground state reactions, for instance changing the population transfer at one of the conical intersections in ethylene derivatives have also been recently developed [57]. Other strong field control approaches include those from Kono et al. [58,59], Kreibich et al. [60] or the use of resonant strong fields, as proposed by Wollenhaupt and Baumert [61].

e) Chirp control

Further development of the control techniques mentioned above resulted in not only controlling the delay time or relative phase of laser pulses, but also their momentary frequency. A laser pulse where the momentary frequency is changed is termed a chirp pulse, in beautiful analogy to the sound waves uttered by birds. If the frequency increases with time, one speaks of an 'up-chirp' and if it decreases, of a 'down-chirp'. Electronic excitations using chirped pulses have been both both theoretically and experimentally investigated; see e.g. refs. [62, 63, 64].

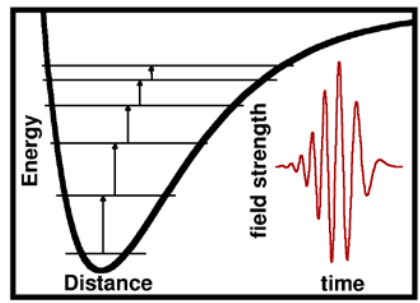

**Fig. 5: Control with chirped pulses.** A chirped pulse, i.e. a laser with the momentary frequency changing, can induce a "ladder climbing" of vibrational states. Here, reducing the laser frequency copes with the subsequently decreasing energy differences of the vibrational "ladder steps".

Another interesting application of chirped pulses is the control of vibrational excitations via ladder-climbing processes (see Fig. 5). Due to anharmonicities, the energy gap between vibrational levels is the smaller the higher is the quantum number. Dissociation by ladder climbing would require as many narrow-band lasers as steps to be resonantly climbed. Alternatively, it is possible to use down-chirped pulses. Both theoretical [65] and experimental [66] research was carried out with these frequency-swept infrared pulses.

Seeking efficient control requires scanning over a set of laser frequencies (and often other parameters). As a generalization, quantum control landscapes (a map where several or possibly all parameters are scanned against the ability to reach a predefined target) were introduced in the field of laser control by Rabitz and co-workers [67] and shortly afterwards recorded experimentally by Cardoza et al. [68] and Wollenhaupt et al. [69]. Such landscapes are a powerful tool to extract not only information on the shape of the underlying potentials but also the velocity, dispersion, and shape of the wave packet [70,71]. The concept is not limited to chirped pulses but a wealth of parameters may be varied, like the intensity or the delay between two pulses.

An exemplary control landscape where excited-state population is analyzed as a function of the spectral phase slope and delay time is shown in Fig. 6.

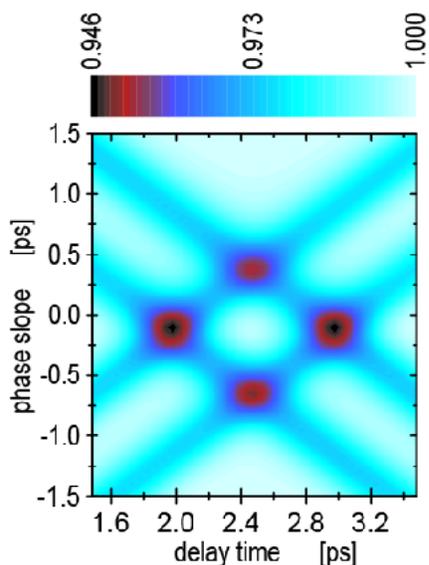

**Fig. 6: Quantum control fitness landscape. Scans of different combined parameter open up a wealth of possibilities for analysis. As an example, the excited state population of a simple model system is plotted against the delay time and the phase slope of a colored double pulse [71].**

If analyzed thoroughly [71], the landscape shows four minima, which can be identified with a wave packet splitting into two fractions entering a designated dumping region with the same velocity and same dispersion.

  f)  Multi-parameter laser control schemes

In the previous strategies analytical laser pulses have been devised where particular parameters like frequency, pulse intensity, temporal length, phase or polarization can be altered to achieve the desired control. In the case of complex systems with many degrees of freedom, the search for adequate analytical pulses is not always easy. First, in larger systems it becomes soon impracticable to obtain global information about the PES. Second, in the presence of many degrees of freedom, the wave packet quickly spreads and leads to complicated distributions, which cannot easily been further transferred to another PES by a subsequent pulse. To circumvent the problem of knowing the PES beforehand, Judson and Rabitz [72] suggested to use genetic algorithms that search the best pulse shapes to prepare specific products based on fitness information, such as product yields. This method, also called adaptive or feedback control, prepares the desired target solving the Schrödinger equation exactly in real time with the true laboratory field. The control system is irradiated with an electric field that prepares a specific product. This output is fed into a computer programmed with a learning algorithm that guides the pulse shaper to produce a new electric field that prepares the desired products in an iterative way. The learning algorithm usually works in the same way as evolution in biology (see Fig. 7). For this reason, the term "genetic algorithm" is employed. A mutation is e.g. introduced by replacing some parameters in a given set defining a control pulse by random numbers. A crossover is carried out by interchanging parameters from different sets.

The first experimental implementation of adaptive control was to transfer population from the ground to the first excited state in a dye [73]. The first chemical application was done in the lab by Gerber and coworkers [74], who starting from $CpFe(CO)_2Cl$ ($Cp=^5\eta\text{-}C_5H_5$) optimized the branching ratio between the parent ion $CpFe(CO)Cl^+$ and the photodissociation product $FeCl^+$. After this breakthrough, many other experiments followed [75,76,77,78,79,80,81], also in liquid phase [82,83], or in biological contexts [84].

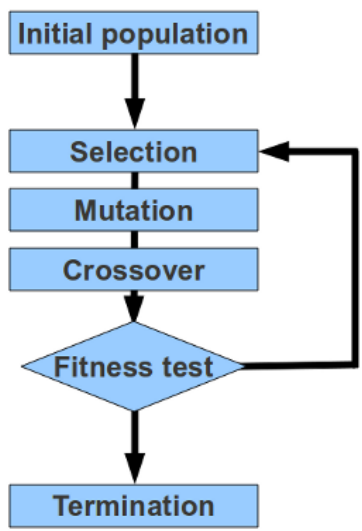

**Fig. 7: Adaptive feedback control. A genetic algorithm is used to modify the parameters of a pulse shaper in an experiment or the laser parameters in a theoretical simulation.**

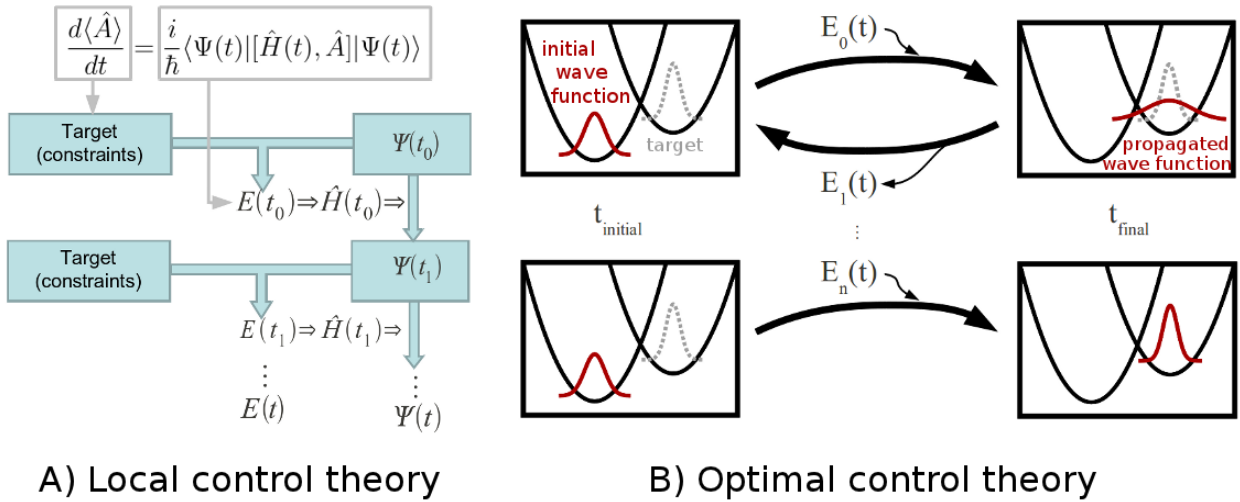

**Fig. 8:** A) Local control theory. The rate of an observable is desired to be positive (or negative or constant). From the rate expression, a formula for the needed laser field can be derived. The laser field drives the dynamics which in turn influences the shape of the field in the next time step. B) Optimal control theory. The method of Lagrangian multipliers is applied to determine the optimal laser field. In practice, several forward and backward propagations are necessary in the dynamical simulation to obtain a converged result.

Adaptive control has also proved successful in controlling nonlinear optical processes, such as the automatic compression of laser pulses [85], two-photon transitions in atoms [86], Rydberg wavepackets [87], generation of high harmonics [88], ultrafast effects in semiconductors [89], and many others.

On the theoretical side, the most popular schemes to find an optimal pulse is to use the so-called optimal control theory (OCT) [28, 90], and local control theory (LCT), but also genetic algorithms can be implemented in purely theoretical simulations [91].

Local control theory (LCT) was invented by Tannor, Kosloff and coworkers [92] and a little later, independently by Rabitz and coworkers, where it is called tracking [93, 94, 95]. A review is given in Ref. 16. In LCT, the control field is determined from the system's dynamics at every instant in time and immediately fed back into the dynamics. Although mainly applied in theoretical simulations, in principle it can be implemented in the laboratory through automated experiments [96]. The laser field is derived from the rate dA/dt of a target expectation value A, where the rate is chosen to remain positive (negative, constant) at all times by the help of the laser field, see Fig. 8A. The field is local in time, since it is determined to achieve a monotonic increase (decrease, constant) in the desired objective. As the field directly stems from the dynamics, it is very close to intuition and in most cases easily understandable. For example, the STIRAP scheme automatically emerged from the local optimization and was extended from three to a system of N levels [97]. Local control was also formulated within the density matrix approach [98,99]. An extension for controlling not only the field's amplitude but also the phase was implemented for quantum computing [100]. Various other applications can be found in the literature (see e.g. references [101, 102, 103]).

In contrast to LCT, OCT searches for maximizing at some specific time a given functional of the radiation field. The formalism employed to optimize this functional is an extension of the variational method where the constraints include differential equations. One constraint is that the amplitude should satisfy the Schrödinger equation, which in practice leads to several forward and backward propagations to iteratively determine the optimal field (see Fig. 8B). Another constraint is that the energy per pulse is limited. The method for finding the extrema of a function of several variables subject to one or more constraints is the method of Lagrange multipliers, which is the basic tool in nonlinear constrained optimizations. From an intuitive point of view, OCT consists of choosing the appropriate wavelengths in each point of the PES such that the wave packet is directed toward the desired channel. Because vibrations take place in fs, the optimally shaped laser pulses are in the scale of fs. Tersigni, Gaspard and Rice employed first OCT to computationally optimize the transfer of population from one state to another [104]. Since then, OCT has proven to be helpful in plenty of applications. Few selected examples include the control of photofragmentation [105], isomerization [106,107,108,109,110], ring-opening reactions [111], photoassociation [112], desorption from metal surfaces [113,114], inducing molecular rotations [115,116], preparation of pure enantiomers [117,118], as well quantum computation [119,120,121] and Bose-Einstein condensates [122].

## 3. From wave packets to classical trajectories.

As mentioned above, large molecules, beyond few atoms, are particularly challenging to describe theoretically because very often several PESs are involved in the photoinduced dynamics, a large amount of degrees of freedom have to be taken into account, and moreover, different types of couplings among the PESs add an extra degree of complexity to the description of photoinduced processes. From the theoretical viewpoint, the application of any of the control strategies explained above requires the knowledge of the PES of the system in advance. In order to solve the time-dependent Schrödinger equation, typically, a pre-calculation of the most important PESs along few degrees of freedom is done. While the calculation of potential energy surfaces is performed mostly routinely for the electronic ground state, this is not the case for excited states. The methods required to calculate excited states are demanding and therefore, it is only possible to calculate accurate PESs in few degrees of freedom, and thus perform wavepacket propagations in reduced dimensionality.

To overcome these limitations, different approximations have been developed. Rather accurate schemes which are able to describe photoinduced dynamics of multidimensional systems include the multi-configurational time-dependent Hartree method [123] or the ab initio multiple spawning method [124,125].

An alternative to employ quantum dynamics is the use of ab initio molecular dynamics (MD). In the latter approach, the equations of motion of the system are divided into two parts: the Newton equations for the nuclei and the time-dependent Schrödinger equation for the electrons. The electronic motion provides the required gradients for the Newton equations and the geometrical parameters obtained in the classical evolution of the nuclei are the input of the quantum calculations. Since the classical motion can be bound to only one state, the Tully surface hopping (SH) algorithm [126] is often used to select the potential and the gradients that govern the classical motion. SH was originally developed to account for nonadiabatic couplings in the photodynamics of molecules. Recent efforts try to describe other types of coupling which may occur during or after photoexcitations. Thachuk et al. treated laser-induced couplings for the first time in SH [127]. Later, also Jones et al. developed another variant using the density matrix approach [128]. The density matrix formalism is also applied in a method developed by Mitrić and coworkers [129], which has been applied in several studies [130,131,132,133]. SH with laser interactions in the framework of density functional theory has been developed by Tavernelli et al. [134]. Spin-orbit coupling within MD has been implemented by Maiti et al. [135]. Lately, a method called SHARC (Surface Hopping in the Adiabatic Representation Including Arbitrary Couplings) able to treat all kind of couplings on the same footing has been developed [136].

To explain the idea behind the SHARC method, it is useful to present the central feature of the original SH scheme. As explained above, SH provides the possibility to "hop" from one potential surface to another. The probability for such a hop is computed from an expression which depends on the Hamiltonian of the system containing the potential and kinetic energy (globally denoted e.g. by a matrix H) and another term that consists of the non-adiabatic or kinetic couplings (denoted e.g. by K). A very important point in the SH scheme is the choice of the basis functions. Tully demonstrated that SH is not invariant with respect to the representation and he recommended working in the adiabatic basis [126]. In this case, the potential matrix contained in H is always diagonal and the coupling between the surfaces is included in the kinetic part K. In quantum dynamics, in contrast, it is desirable to work in the diabatic regime since then the K matrix is strictly zero and the coupling between the different electronic basis functions is included in the off-diagonal elements of the potential matrix of H. Most *ab initio* quantum chemistry programs yield potential energies in the adiabatic representation, i.e. yield adiabatic potentials and non-adiabatic or kinetic couplings. However, as e.g. spin-orbit coupling or the electric-field interaction entering through the dipole moment matrix (permanent and transition ones) are not included when obtaining the electronic wave functions, such couplings are usually incorporated *a posteriori* in the potential part of the Hamiltonian. As a result, the resulting Hamiltonian matrix cannot be called adiabatic anymore, even if the so-called "adiabatic" potential surfaces are employed.

In order to include the effect of the spin-orbit coupling and the electric field interaction in the classical motion, SHARC "fully" adiabatizes the Hamiltonian matrix H with respect to any arising coupling via a unitary transformation. The hopping probability is modified by this transformation and then the description of laser interactions and spin-orbit coupling effects is possible.

SHARC has been applied to model the laser-induced dynamics of the IBr molecule [136] and also to simulate effects of Stark control [137]. After photoexcitation, IBr dissociates into two channels, I+Br and I+Br* (see Fig. 9A). Although the system seems rather simple at first sight, it is a real challenge for theoretical simulations. Spin-orbit coupling leads on the one hand to a very anharmonic ground state potential.

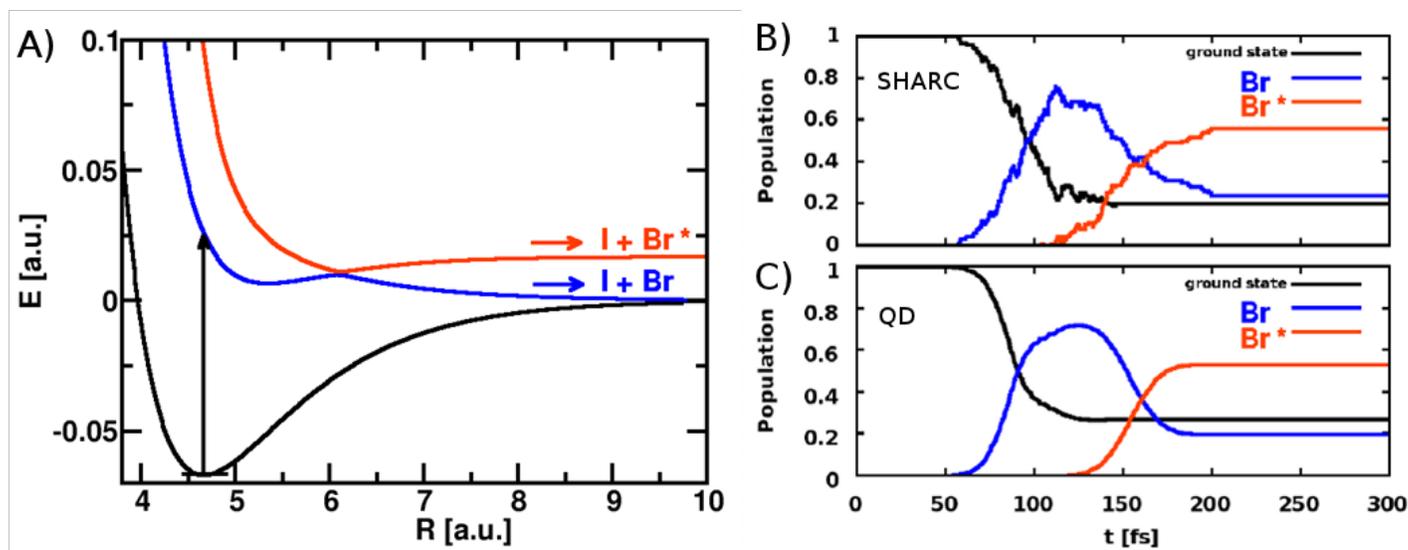

**Fig. 9.** From quantum to molecular dynamics. A: Potential energy curves for the photo-dissociation of the IBr molecule. B: Population dynamics after photoexcitation with a laser pulse centered at 100 fs as calculated from the molecular dynamics SHARC approach. C) Exact quantum dynamical (QD) results for comparison. Adapted from [136].

On the other hand, the main features of the excited state dynamics are neither described by a single pathway in the purely adiabatic nor in the purely diabatic picture [138], which is the worst-case scenario for the two involved channels. Moreover, a very large distribution of momenta is obtained after photoexcitation due to the extreme steepness of the excited state potentials in the Franck-Condon region. Despite these difficulties, which are not easily treatable within a semiclassical frame, SHARC is able to describe the dissociation dynamics correctly [136] and yields the same branching ratio for the two product channels as an exact quantum dynamical simulation would do (see Fig. 9B,C), in agreement with the experiment.

After the IBr molecule is electronically excited with a resonant laser, the natural branching ratio in the two channels can be influenced with a second, nonresonant laser. As described above, the NRDSE is able to change the potentials and in this way acts as a photonic catalyst. In addition, this scenario is extremely challenging for semiclassical computations because the involved intense laser fields modify the potential energy curves in time. It is encouraging that the results obtained from the SHARC method are also in very good agreement with the outcome of exact quantum mechanical simulations [137]. SHARC appears to be a promising method for treating the photoinduced dynamics and control of molecular systems in the presence in any type of coupling taking into account all the degrees of freedom.

## 4. Conclusions and perspectives

In this paper, we have reviewed some of the most important developments in the field of laser control, where our group has made substantial contributions. This field has seen an amazing development in the last decades. What started from simple schemes where only one parameter is changed has evolved to complex control algorithms. Yet, there is plenty of space for further progress.

Despite of the enormous progress, we are still far from the ultimate goal to transform any (waste) substance into another desired product at the chemical level with the same efficiency and precision as traditional chemistry does. In this direction, a particular challenge theoreticians have to face is to describe control in the liquid phase. Most of chemistry takes place in solution. In this respect, the possibility to merge quantum control techniques within ab initio MD combined with molecular mechanical force fields offers a tractable pathway. First steps in this direction are already undertaken [139]. Semiclassical simulations also offer a rather easy way to incorporate the orientational degree of freedom, which is often neglected in quantum dynamics studies.

With the advent of laser control, the shortest pulses were in the femtosecond regime. Consequently, the studies focused mainly on the atomic motion where a vibration typically has a comparable time scale. With the development of attosecond lasers, the realm of electron control becomes accessible and this century will witness large progress in this direction [140].

One of the major challenges remains since the beginning of laser control: As many optimal solutions come from black box strategies – not in the sense that they are easy to use but in the sense that the mechanisms behind the resulting pulses remain unclear - a thorough analysis is necessary. Therefore, theory is indispensable and will even gain in importance in the future.

**Acknowledgements**
The authors would like to thank past and present students for their contributions to the group in the field of laser control. Especially, we would like to thank Martin Richter, Jesus González-Vázquez and Ignacio Sola for their current efforts in SHARC. Financial support from the Deutsche Forschungsgemeinschaft (SFB 450 "Analysis and Control of Ultrafast Photoinduced Reactions" and the project GO 1059/6-1) and German Federal Ministry of Education and Research within the research initiative PhoNa is also gratefully acknowledged.

# References

1 A. H. Zewail, Femtochemistry: atomic-scale dynamics of the chemical bond using ultrafast lasers (Nobel Lecture), *Angew. Chem. Int. Ed.* **39** (2000) 2587-2631.

2 P. Brumer, M. Shapiro, Laser control of molecular processes, *Ann. Rev. Phys. Chem.*, **43** (1992) 257-282.

3 R. J. Gordon, S. A. Rice, Active Control of the dynamics of atoms and molecules, *Ann. Rev. Phys. Chem.*, **48** (1997) 601-641.

4 S. A. Rice, Perspectives on the control of quantum many-body dynamics: Applications to chemical reactions, *Adv. Chem. Phys.*, **101** (1997) 213-283.

5 D. J. Tannor, R. Kosloff, A. Bartana, Laser cooling of internal degrees of freedom of molecules by dynamically trapped states, *Faraday Discuss.*, **113** (1999) 365-383.

6 M. Dantus, Ultrafast four-wave mixing in the gas phase, *Ann. Rev. Phys. Chem.*, **52** (2001) 639-679.

7 T. Brixner, N. H. Damrauer, G. Gerber, Femtosecond quantum control, *Adv. At. Molec. Opt. Phys.*, **46** (2001) 1-54.

8 N. E. Henriksen, Laser control of chemical reactions, *Chem. Soc. Rev.*, **31** (2002) 37-42.

9 T. Brixner, G. Gerber, Quantum control of gas-phase and liquid-phase femtochemistry, *Chem. Phys. Chem.*, **4** (2003) 418-438.

10 M. Shapiro, P. Brumer, Coherent control of molecular dynamics, *Rep. Prog. Phys.*, 66 (2003) 859-942.

11 M. Dantus, A. H. Zewail (Eds.), Special issue on Femtochemistry, *Chem. Rev.*, **104** (2004) 1717-2124.

12 I. V. Hertel, W. Radloff, Ultrafast dynamics in isolated molecules and molecular clusters, *Rep. Prog. Phys.*, **69** (2006) 1897-2003.

13 O. Kühn, L. Wöste (Eds.), Analysis and Control of Ultrafast Photoinduced Reactions, Springer, Berlin, ISBN 0172-6218 (2006).

14 B. Lasorne, G. A. Worth, Coherent Control of Molecules, CCP6, Daresbury Laboratory, ISBN 0-9545289-5-6 (2006).

15 P. Nuernberger, G. Vogt, T. Brixner, G. Gerber, Femtosecond quantum control of molecular dynamics in the condensed phase, *Phys. Chem. Chem. Phys.*, **9** (2007) 2470-2497.

16 V. Engel, C. Meier, D. Tannor, Local control theory: recent applications to energy and particle transfer processes in molecules, *Adv. Chem. Phys.*, **141** (2009) 29-101.

17 G. A. Worth, C. Sanz-Sanz, Guiding the time-evolution of a molecule: optical control by computer, *Phys. Chem. Chem. Phys.*, 12 (2010) 15570-15579.

18 J. Manz, L. Wöste (Eds), Femtosecond Chemistry, VCH, ISBN 978-3527290628 (1995).

19 R. N. Zare, Laser control of chemical reactions, *Science,* 279 (1998) 1875-1879.

20 C. J. Bardeen, Q. Wang, C. V. Shank, Selective excitation of vibrational wave-packet motion using chirped pulses, *Phys. Rev. Lett.,* **75** (1995) 3410-3413.

21 C. J. Bardeen, V. V. Yakovlev, J. A. Squier, K. R. Wilson, Quantum control of population transfer in green fluorescent protein by using chirped femtosecond pulses, *J. Am. Chem. Soc.*,**120** (1998) 13023-13027.

22 P. Brumer and M. Shapiro, Control of unimolecular reactions using coherent-light, *Chem. Phys. Lett.* **126** (1986**)** 541-546**.**

23 M. Shapiro, J. W. Hepburn and P. Brumer, Simplified laser control of unimolecular reactions: Simultaneous ($\omega_1$, $\omega_3$) excitation, *Chem. Phys. Lett.*, **149** (1988) 451-454.

24 M. Shapiro, J. W. Hepburn, P. Brumer, Simplified laser control of unimolecular reactions: simultaneous ($\omega_1$, $\omega_3$) excitation, *Chem. Phys. Lett.*, **149** (1988) 451–454.

25 C. K. Chan, P. Brumer, M. Shapiro, Coherent radiative control of IBr photodissociation via simultaneous ($\omega_1$,$\omega_3$) excitation, *J. Chem. Phys.,* **94** (1991) 2688-2696.

26 C. Chen, Y. Y. Yin, D. S. Elliott, Interference between optical-transitions, *Phys. Rev. Lett.* **64** (1990) 507-510.

27 L. C. Zhu, V. D. Kleiman, X. N. Li, S. P. Lu, K. Trentelman, R. J Gordon, Coherent laser control of the product distribution obtained in the photoexcitation of HI, *Science*, **270** (1995) 77-80.

28 D. J. Tannor, S. A. Rice, Control of selectivity of chemical reactions via control of wave packet evolution, *J. Chem. Phys.*, **83** (1985) 5013-5018.

29 D. J. Tannor, R. Kosloff, S. A. Rice, Coherent pulse sequence induced control of selectivity of reactions - exact quantum-mechanical calculations, *J. Chem. Phys.*, **85** (1986) 5805-5820.

30 R. J. Gordon, S. A. Rice, Active control of the dynamics of atoms and molecules, *Annu. Rev. Phys. Chem.*, **48** (1997) 601-641.

31 T. Baumert, B. Buhler, R. Thalweiser, G. Gerber, Femtosecond spectroscopy of molecular autoionization and fragmentation, *Phys. Rev. Lett.*, **64** (1990) 733-736.

32 T. Baumert, M. Grosser, R. Thalweiser, G. Gerber, Femtosecond time-resolved molecular multiphoton ionization - the $Na_2$ system, *Phys. Rev. Lett.*, **67** (1991) 3753-3756.

33 E. D. Potter, J. L. Herek, S. Pedersen, Q. Liu, A. H. Zewail, Femtosecond laser control of a chemical-reaction, *Nature*, **355** (1992) 66-68.

34 B. Amstrup, N. E. Henriksen, Two-pulse laser control of bond-selective fragmentation, *J. Chem. Phys* , **105** (1996) 9115-9120.

35 B. Amstrup and N. E. Henriksen, Control of HOD photodissociation dynamics via bond-selective infrared multiphoton excitation and a femtosecond ultraviolet laser pulse, *J. Chem. Phys.*, **97** (1992) 8285-8295.

36  N. E. Henriksen, Theoretical concepts in molecular photodissociation dynamics, *Adv. Chem. Phys.*, **91** (1995) 433-509.

37  N. Elghobashi, L. González, J. Manz, Quantum model simulations of symmetry breaking and control of bond selective dissociation of FHF- using IR+UV laser pulses, *J. Chem. Phys.*, **120** (2004) 8002-8014.

38  N. Elghobashi and J. Manz, Separating the photofragments of randomly oriented symmetric reactants by IR+UV laser pulses: Quantum simulations for FHF--> F+FH+e, *Isr. J. Chem.,* **43** (2003) 293-303.

39  N. Elghobashi, L. González, J. Manz, Quantum Simulations for isotope effects of IR + UV laser pulses on symmetry and selective hydrogen bond breaking, *Z. Phys. Chem.*, **217** (2003) 1577-1596.

40  N. Elghobashi, L. González, J. Manz, Quantum model simulations of symmetry breaking and control of bond selective dissociation of FHF- using IR+UV laser pulses, *J. Chem. Phys.*, **120** (2004) 8002-8014.

41  T. Rozgonyi, L. González, Control of the photodissociation of CH2BrCl using a few-cyle IR driving laser pulse and a UV control pulse, *Chem. Phys. Lett.*, **459** (2008) 39-43.

42  N. Elghobashi, P. Krause, J. Manz and M. Oppel, IR+UV laser pulse control of momenta directed to specific products: Quantum model simulations for HOD*-> H+OD versus HO+D, *Phys. Chem. Chem. Phys.*, **5** (2003) 4806-4813.

43  Y. Fujimura, L. González, D. Kröner, J. Manz, I. Mehdaoui, B. Schmidt, Quantum ignition of intramolecular rotation by means of IR + UV laser pulses, *Chem. Phys. Lett.*, **386** (2004) 248-253.

44  G. Pérez-Hernández, A. Pelzer, L. González, T. Seideman, Biologically-Inspired Molecular Machines Driven by Light. Optimal Control of a Unidirectional Rotor, *New J. Phys.*, **12** (2010) 75007.

45  K. Hoki, M. Sato, M. Yamaki, R. Sahnoun, L. González, S. Koseki, Y. Fujimura, Chiral Molecular Motors Ignited by Femtosecond Pump-Dump Laser Pulses, *J. Phys. Chem. B*, **108** (2004) 4916-4921.

46  U. Gaubatz, P. Rudecki, M. Becker, S. Schiemann, M. Külz, K. Bergmann, Population switching between vibrational levels in molecular-beams, *Chem. Phys. Lett.*, **149** (1988) 463-468.

47  K. Bergmann, H. Theuer and B. W. Shore, Coherent population transfer among quantum states of atoms and molecules, *Rev. Mod. Phys.*, **70** (1998) 1003-1025.

48  K. Bergmann and B. W. Shore, in: Molecular Dynamics and Spectroscopy by Stimulated Emission Pumping, World Scientific Publishing Company, ISBN 978-9810221119 (1995).

49  K. Bergmann, H. Theuer and B. W. Shore, Coherent population transfer among quantum states of atoms and molecules, *Rev. Mod. Phys.*, **70** (1998) 1003-1025.

50  T. Halfmann, K. Bergmann, Coherent population transfer and dark resonances in $SO_2$, *J. Chem. Phys.*, **104** (1996) 7068-7072.

51  L. González, D. Kröner, I.R. Solá, Separation of Enantiomers by UV Laser Pulses in H2POSH: π-Pulses vs. Adiabatic Transitions, *J. Chem. Phys.*, **115** (2001) 2519-2529.

52  I. Thanopulos, M. Shapiro, Detection and Automatic Repair of Nucleotide Base-Pair Mutations by Coherent Light, *J. Am. Chem. Soc.*, **127** (2005) 14434-14438.

53  D. Geissler, T. Rozgonyi, J. González-Vázquez, L. González, S. Nichols, T. Weinacht, Creation of Multi-Hole Molecular Wave Packets via Strong Field Ionization, *Phys. Rev. A*, **82** (2010) 011402.

54  J. González-Vázquez, I. R. Sola, J. Santamaría, Adiabatic passage by light-induced potentials in polyatomic molecules, *J. Phys. Chem. A*, **110** (2006) 1586-1593.

55  B. J. Sussman, D. Townsend, M. Y. Ivanov, A. Stolow, Dynamic stark control of photochemical processes, Science, **314** (2006) 278-281.

56  J. González-Vázquez, I. R. Sola, J. Santamaría, V. S. Malinovsky, Optical control of the singlet-triplet transition in $Rb_2$, *J. Chem. Phys.*, **125** (2006) 124315.

57  J. González-Vázquez, I. R. Sola, J. Santamaría, L. González, unpublished results (2008).

58  H. Kono, Y. Sato, N. Tanaka, T. Kato, K. Nakai, S. Koseki, Y. Fujimura, Quantum mechanical study of electronic and nuclear dynamics of molecules in intense laser fields, *Chem. Phys.*, **304** (2004) 203-226.

59  M. Kanno, T. Kato, H. Kono, Y. Fujimura, F. H. M. Faisal, Incorporation of a wave-packet propagation method into the S-matrix framework: Investigation of the effects of excited state dynamics on intense-field ionization, *Phys. Rev. A*, **72** (2005) 033418.

60  T. Kreibich, R.van Leeuwen, E. K. U. Gross, Time-dependent variational approach to molecules in strong laser fields, *Chem. Phys.*, **304** (2004) 183-202.

61  M. Wollenhaupt, T. Baumert, Ultrafast strong field quantum control on $K_2$ dimers, *J. Photochem. Photobio. A: Chem.*, **180** (2006) 248-255.

62  B. Kohler, V. V. Yakovlev, J. Che, J. L. Krause, M. Messina, K. R. Wilson, N. Schwentner, R. M. Whitnell, Y. Yan, Quantum Control of Wave Packet Evolution with Tailored Femtosecond Pulses, *Phys. Rev. Lett.*, **74** (1995) 3360-3363.

63  T. Lohmüller, M. Erdmann, V. Engel, Chirped pulse ionization: bondlength dynamics and interference effects, *Chem. Phys. Lett.*, **373** (2003) 319–327.

64  C. J. Bardeen, Q. Wang, C. V. Shank, Selective Excitation of Vibrational Wave Packet Motion Using Chirped Pulses, *Phys. Rev. Lett.*, **75** (1995) 3410–3413.

65  S. Chelkowski, A. D. Bandrauk, P. B. Corkum, Efficient molecular dissociation by a chirped ultrashort laser pulse, *Phys. Rev. Lett.*, **65** (1990) 2355–2358.

66  B. Broers, H. B. van Linden van den Heuvell, L. D. Noordam, Efficient population transfer in a three-level

ladder system by frequency-swept ultrashort laser pulses, *Phys. Rev. Lett.*, **69** (1992) 2062–2065.

67   H. Rabitz, M. Hsieh, C. Rosenthal, Quantum Optimally Controlled Transition Landscapes, *Science*, **303** (2004) 1998-2001.

68   D. Cardoza, C. Trallero-Herrero, F. Langhojer, H. Rabitz, T. Weinacht, *J. Chem. Phys.*, **122** (2005) 124306.

69   M. Wollenhaupt, A. Präkelt, C. Sarpe-Tudoran, D. Liese, and T. Baumert, *J. Mod. Opt.*, **52** (2005) 2187-2195.

70   P. Marquetand, P. Nuernberger, G. Vogt, T. Brixner, V. Engel, Properties of wave packets deduced from quantum control fitness landscapes, *Europhys. Lett.*, **80** (2007) 53001.

71   P. Marquetand, P. Nuernberger, T. Brixner, V. Engel, Molecular dump processes induced by chirped laser pulses, *J. Chem. Phys.*, **129** (2008) 074303.

72   R. S. Judson, H. Rabitz, Teaching lasers to control molecules, *Phys. Rev. Lett.*, **68** (1992) 1500-1503.

73   C. J. Bardeen, V. V. Yakovlev, K .R. Wilson, S. D. Carpenter, P. M. Weber, W. S. Warren, Feedback quantum control of molecular electronic population transfer, *Chem. Phys. Lett.*, **280** (1997) 151-158.

74   A. Assion, T. Baumert, M. Bergt, T. Brixner, B. Kiefer, V. Seyfried, M. Strehle, G. Gerber, Control of chemical reactions by feedback-optimized phase-shaped femtosecond laser pulses, *Science*, **282** (1998) 919-922.

75   A. Glass, T. Rozgonyi, T. Feurer, G. Szabó, Control of the photodissociation of CsCl, *Appl. Phys. B*, **71** (2000) 267-276.

76   T. Feurer, A. Glass, T. Rozgonyi, G. Szabó, Control of the photodissociation process of CsCl using a feedback-controlled self-learning fs-laser system, *Chem. Phys.*, **267** (2001) 223-229.

77   S. Vajda, P. Rosendo-Francisco, C. Kaposta, M. Krenz, C. Lupulescu, L. Wöste, Analysis and control of ultrafast photodissociation processes in organometallic molecules, *Eur. Phys. J. D*, **16** (2001) 161-164.

78   C. Daniel, J. Full, L. González, C. Kaposta, M. Krenz, C. Lupulescu, J. Manz, S. Minemoto, M. Oppel, P. R. Francisco, S. Vajda, L. Wöste, Analysis and control of laser induced fragmentation processes in CpMn(CO)(3), *Chem. Phys.*, **267** (2001) 247-260.

79   C. Daniel, J. Full, L. González, C. Lupulescu, J. Manz, A. Merli, S. Vajda, L. Wöste, Deciphering the reaction dynamics underlying optimal control laser fields, *Science*, **299** (2003) 536-539.

80   N. H. Damrauer, C. Dietl, G. Krampert, S. H. Lee, K. H. Jung, G. Gerber, Control of bond-selective photochemistry in CH2BrCl using adaptive femtosecond pulse shaping, *Eur. Phys. J. D*, **20** (2003) 71-76.

81   T. Brixner, N. H. Damrauer, G. Krampert, P. Niklaus, G. Gerber, Femtosecond learning control of quantum dynamics in gases and liquids: technology and applications, *J. Mod. Opt.*, **50** (2003) 539-560.

82   T. Brixner, N. H. Damrauer, P. Niklaus, G. Gerber, Photoselective adaptive femtosecond quantum control in the liquid phase, *Nature*, **414** (2001) 57-60.

83   T. Brixner, G. Gerber, Quantum control of gas-phase and liquid-phase femtochemistry, *Chem. Phys. Chem.*, **4** (2003) 418-438.

84   J. L. Herek, W. Wohlleben, R. J. Cogdell, D. Zeidler, M. Motzkus, Quantum control of energy flow in light harvesting, *Nature*, **417** (2002) 533-535.

85   D. Yelin, D. Meshulach, Y. Silberberg, Adaptive femtosecond pulse compression, *Opt. Lett.*, **22** (1997) 1793-1795.

86   D. Meshulach, Y. Silberberg, Coherent quantum control of two-photon transitions by a femtosecond laser pulse, *Nature*, **396** (1998) 239-242.

87   T. C. Weinacht, J. Ahn, P. H. Bucksbaum, Controlling the shape of a quantum wavefunction, *Nature*, **397** (1999) 233-235.

88   R. Bartels, S. Backus, E. Zeek, L. Misoguti, G. Vdovin, I. P. Christov, M. M. Murnane, H. C. Kapteyn, Shaped-pulse optimization of coherent emission of high-harmonic soft X-rays, *Nature*, **406** (2000) 164-166.

89   J. Kunde, B. Baumann, S. Arlt, F. Morier-Genoud, U. Siegner, U. Keller, Adaptive feedback control of ultrafast semiconductor nonlinearities, *Appl. Phys. Lett.*, **77** (2000) 924-926.

90   R. Kosloff, S. A. Rice, P. Gaspard, S. Tersigni, D. J. Tannor, Wavepacket dancing - achieving chemical selectivity by shaping light-pulses, *Chem. Phys.*, **139** (1989) 201-220.

91   W. Zhu, H. Rabitz , Closed loop learning control to suppress the effects of quantum decoherence, *J. Chem. Phys.*, **118** (2003) 6751-6757.

92   R. Kosloff, A. D. Hammerich, D. J. Tannor, Excitation without demolition: Radiative excitation of ground-surface vibration by impulseive stimulated Raman scattering with damage control, *Phys. Rev. Lett.*, **69** (1992) 2172–2175.

93   P. Gross, H. Singh, H. Rabitz, K. Mease, G. M. Huang, Inverse quantummechanical control: A means for design and a test of intuition, *Phys. Rev. A*, **47** (1993) 4593–4604.

94   Y. Chen, P. Gross, V. Ramakrishna, H. Rabitz, K. Mease, H. Singh, Control of classical regime molecular objectives – applications of tracking and variations on the theme, *Automatica*, **33** (1997) 1617–1633.

95   W. Zhu, H. Rabitz, Quantum control design via adaptive tracking, *J. Chem. Phys.*, **119** (2003) 3619–3625.

96   D. J. Tannor, R. Kosloff, A. Bartana, Laser cooling of internal degrees of freedom of molecules by dynamically trapped states, *Faraday Discuss.*, **113** (1999) 365–383.

97   V. Malinovsky, D. J. Tannor, Simple and robust extension of the stimulated Raman adiabatic passage technique to N-level systems, *Phys. Rev. A*, **56** (1997) 4929–4937.

98  A. Bartana, R. Kosloff, D. J. Tannor, Laser cooling of molecular internal degrees of freedom by a series of shaped pulse, *J. Chem. Phys.*, **99** (1993) 196–210.

99  H. Tang, R. Kosloff, S. A. Rice, A generalized approach to the control of the evolution of a molecular system, *J. Chem. Phys.*, **104** (1996) 5457–5471.

100  S. E. Sklarz, D. J. Tannor, Quantum computation via local control theory: direct sum vs. direct product Hilbert spaces, *Chem. Phys.*, **322** (2006) 87–97.

101  S. Gräfe, P. Marquetand, N. E. Henriksen, K. B. Møller, V. Engel, Quantum control fields from instantaneous dynamics, *Chem. Phys. Lett.*, **398** (2004) 180–185.

102  C. Meier, M.-C. Heitz, Laser control of vibrational excitation in carboxyhemoglobin: A quantum wave packet study, *J. Chem. Phys.*, **123** (2005) 044504.

103  M. Sugawara, Y. Fujimura, Control of quantum dynamics by a locally optimized laser. Application to ring puckering isomerization, *J. Chem. Phys.*, **100** (1994) 5646–5655.

104  S. H. Tersigni, P. Gaspard, S. A. Rice, On using shaped light-pulses to control the selectivity of product formation in a chemical-reaction - an application to a multiple level system, *J. Chem. Phys.,* **93** (1990) 1670-1680.

105  M. Abe, Y. Ohtsuki, Y. Fujimura, Z. Lan, W. Domcke, Geometric phase effects in the coherent control of the branching ratio of photodissociation products of phenol, *J. Chem. Phys.*, **124** (2006) 224316.

106  J. Manz, K. Sundermann, R. de Vivie-Riedle, Quantum optimal control strategies for photoisomerization via electronically excited states, *Chem. Phys. Lett.*, **290** (1998) 415-422.

107  R. Mitrić, M. Hartmann, J. Pittner, V. Bonačić-Koutecký, New strategy for optimal control of femtosecond pump-dump processes, *J. Phys. Chem. A*, **106** (2002) 10477.

108  M. Artamonov, T.-S. Ho, H. Rabitz, Quantum optimal control of ozone isomerization, *Chem. Phys.*, **305** (2004) 213-222.

109  M. Abe, Y. Ohtsuki, Y. Fujimura, W. Domcke, Optimal control of ultrafast cis-trans photoisomerization of retinal in rhodopsin via a conical intersection, *J. Chem. Phys.*, **123** (2005) 144508.

110  V. Bonačić-Koutecký, R. Mitrić, Theoretical Exploration of Ultrafast Dynamics in Atomic Clusters; Analysis and Control, *Chem. Rev.*, **105** (2005) 11-65.

111  D. Geppert, R. de Vivie-Riedle, Control strategies for reactive processes involving vibrationally hot product status, *J. Photochem. Photobiol.*, **180** (2006) 282-288.

112  C. P. Koch, J. P. Palao, R. Kosloff, F. Masnou-Seeuws, Stabilization of ultracold molecules using optimal control theory, *Phys. Rev. A*, **70** (2004) 013402.

113  K. Nakagami, Y. Fujimura, Hybrid quantum control of photodesorption of NO from a metal surface, *Chem. Phys. Lett.*, **360** (2002) 91-98.

114  P. Saalfrank, M. Nest, I. Andrianov, T. Klamroth, D. Kröner, S. Beyvers, Quantum dynamics of laser-induced desorption from metal and semiconductor surfaces, and related phenomena, *J. Phys.: Condens. Matter,* **18** (2006) S1425-S1459.

115  K. Hoki, M. Sato, M. Yamaki, R. Sahnoun, L. González, S. Koseki, Y.Fujimura, Chiral molecular motors ignited by femtosecond pump-dump laser pulses, *J. Phys. Chem. B*, **108** (2004) 4916-4921.

116  G. Pérez-Hernández, A. Pelzer, L. González, T. Seideman, Biologically-Inspired Molecular Machines Driven by Light. Optimal Control of a Unidirectional Rotor, *New J. Phys.*, **12** (2010) 075007.

117  L. González, K. Hoki, D. Kröner, A. S. Léal, J. Manz, Y. Ohtsuki, Selective preparation of enantiomers by laser pulses: From optimal control to specific pump and dump transitions, *J. Chem. Phys.*, **113** (2000) 11134-11142.

118  K. Hoki, S. Koseki, T. Matsushita, R. Sahnoun, Y. Fujimura, Quantum control of molecular chirality: Ab initio molecular orbital study and wave packet analysis of 1,1 '-binaphthyl, *J. Photochem. Photobio. A: Chem.*, **178** (2006) 258-263.

119  C. Tesch, R. de Vivie-Riedle, Quantum Computation with Vibrationally Excited Molecules, *Phys. Rev. Lett.*, **89** (2002) 157901.

120  S. E. Sklarz, D. J. Tannor, Quantum computation via local control theory: Direct sum vs. direct product Hilbert spaces, *Chem. Phys.*, **322** (2006) 87-97.

121  R. de Vivie-Riedle, U. Troppmann, Femtosecond lasers for quantum information technology, *Chem. Rev.*, **107** (2007) 5082-5100.

122  T. Hornung, S. Gordienko, R. de Vivie-Riedle, B. J. Verhaar, Optimal conversion of an atomic to a molecular Bose-Einstein condensate, *Phys. Rev. A*, **66** (2002) 043607.

123  H.-D. Meyer, F. Gatti, G. A. Worth, Eds. Multidimensional Quantum Dynamics: MCTDH Theory and Applications. Wiley-VCH, Weinheim (2009).

124  B. G. Levine, J.D. Coe, A. M. Virshup, T. J. Martínez, Implementation of ab initio multiple spawning in the Molpro quantum chemistry package, *Chem. Phys.*, **347** (2008) 3-16.

125  A. M. Virshup, C. Punwong, T. V. Pogorelov, B. A. Lindquist, C. Ko, T. J. Martínez, Photodynamics in Complex Environments: Ab Initio Multiple Spawning Quantum Mechanical/Molecular Mechanical Dynamics, *J. Phys. Chem. B*, **113** (2009) 3280-3291.

126  J. C. Tully, Molecular dynamics with electronic transitions, *J. Chem. Phys.*, **93** (1990) 1061-1071.

127  M. Thachuk, M. Y. Ivanov, D. M. Wardlaw, A semiclassical approach to intense-field above-threshold dissociation in the long wavelength limit, *J. Chem. Phys.*, **105** (1996) 4094-4104.

128  G. A. Jones, A. Acocella, F. Zerbetto, On-the-Fly, Electric-Field-Driven, Coupled Electron−Nuclear Dynamics, *J. Phys. Chem. A*, **112** (2008) 9650–9656.

[129] R. Mitrić, J. Petersen, V. Bonačić-Koutecký, Laser-field-induced surface-hopping method for the simulation and control of ultrafast photodynamics, *Phys. Rev. A*, **79** (2009) 053416.

[130] J. Petersen, R. Mitrić, V. Bonačić-Koutecký, J.-P. Wolf, J. Roslund, H. Rabitz, How Shaped Light Discriminates Nearly Identical Biochromophores, *Phys. Rev. Lett.*, **105** (2010) 073003.

[131] R. Mitrić, J. Petersen, M. Wohlgemuth, U. Werner, V. Bonačić-Koutecký, L. Wöste, J. Jortner, Time-Resolved Femtosecond Photoelectron Spectroscopy by Field-Induced Surface Hopping, *J. Phys. Chem. A*, **115** (2011) 3755-3765.

[132] R. Mitrić, J. Petersen, M. Wohlgemuth, U. Werner, V. Bonačić-Koutecký, Field-induced surface hopping method for probing transition state nonadiabatic dynamics of $Ag_3$, *Phys. Chem. Chem. Phys.*, **13** (2011) 8690-8696.

[133] P. Lisinetskaya, R. Mitrić, Simulation of laser-induced coupled electron-nuclear dynamics and time-resolved harmonic spectra in complex systems, *Phys. Rev. A*, 83 (2011) 033408.

[134] I. Tavernelli, B. F. E. Curchod, U. Rothlisberger, Mixed quantum-classical dynamics with time-dependent external fields: A time-dependent density-functional-theory approach, *Phys. Rev. A*, **81** (2010) 052508.

[135] B. Maiti, G. C. Schatz, G. Lendvay, Importance of Intersystem Crossing in the S(3P, 1D) + $H_2$ → SH + H Reaction, *J. Phys. Chem. A*, **108** (2004) 8772-8781.

[136] M. Richter, P. Marquetand, J. González-Vázquez, I. Sola, L. González, SHARC - ab initio molecular dynamics with surface hopping in the adiabatic representation including arbitrary couplings, *J. Chem. Theory Comput.*, **7** (2011) 1253-1258.

[137] P. Marquetand, M. Richter, J. González-Vázquez, I. Sola, L. González
Nonadiabatic ab initio molecular dynamics including spin-orbit coupling and laser fields
*Faraday Discuss.*, **153** (2011) 261-273.

[138] M. Shapiro, M. J. J. Vrakking, A. Stolow, Nonadiabatic wave packet dynamics: Experiment and theory in IBr, *J. Chem. Phys.,* **110** (1999) 2465-2473.

[139] I. Tavernelli, B. F.E. Curchod, U. Röthlisberger, Nonadiabatic molecular dynamics with solvent effects: A LR-TDDFT QM/MM study of ruthenium (II) tris (bipyridine) in water, *Chem. Phys.*, **391** (2011) 101-109.

[140] M. F. Kling, M. J. J. Vrakking, Attosecond Electron Dynamics, *Ann. Rev. Phys. Chem.*, **59** (2008) 463-492.